\documentclass[twocolumn,showpacs,amsmath,amssymb,aps,showkeys,floatfix,prd,a4paper]{revtex4}

\usepackage[dvips]{graphicx}
\usepackage{dcolumn}
\usepackage{bm}
\usepackage{epsfig}
\usepackage{amsfonts}
\usepackage{amssymb,amscd}

\def\lsim{\raise0.3ex\hbox{$<$\kern-0.75em\raise-1.1ex\hbox{$\sim$}}}
\def\gsim{\raise0.3ex\hbox{$>$\kern-0.75em\raise-1.1ex\hbox{$\sim$}}}

\def\odd{{I\!\!\!O}}
\def\beq{\begin{equation}}
\def\eeq{\end{equation}}
\def\bea{\begin{eqnarray}}
\def\eea{\end{eqnarray}}
\def\bq{\begin{quote}}
\def\eq{\end{quote}}

\def\gappeq{\mathrel{\rlap {\raise.5ex\hbox{$>$}}
{\lower.5ex\hbox{$\sim$}}}}

\def\lappeq{\mathrel{\rlap{\raise.5ex\hbox{$<$}}
{\lower.5ex\hbox{$\sim$}}}}

\def\Toprel#1\over#2{\mathrel{\mathop{#2}\limits^{#1}}}

\begin{document}


\title{Probing the  Odderon  in  coherent hadron - hadron interactions at CERN LHC}

\author{V.~P. Gon\c{c}alves}
\email{barros@ufpel.edu.br}
\affiliation{High and Medium Energy Group, \\
Instituto de F\'{\i}sica e Matem\'atica, Universidade Federal de Pelotas\\
Caixa Postal 354, CEP 96010-900, Pelotas, RS, Brazil}
\date{\today}

\begin{abstract}

One of the open questions of the strong interaction theory is 
the existence of the Odderon, which is an unambiguous prediction of Quantum Chromodynamics, but still not confirmed in the experiment. In this paper we propose the study of the diffractive $\eta_c$ photoproduction in coherent interactions as  a new alternative to probe the Odderon in $pp$ and $PbPb$ collisions at CERN - LHC.
As the Pomeron exchange cannot contribute to this process, its  observation  would indicate the existence of the Odderon. We predict  total cross sections of order of $pb \, (\mu b)$ for $pp \, (PbPb)$ collisions and large values for the event rates/year, which makes, in principle, the experimental analysis of this process feasible at LHC.
\end{abstract}

\pacs{12.38.Aw, 13.85.Lg, 13.85.Ni}
\keywords{Quantum Chromodynamics, Meson production,  Coherent interactions}

\maketitle

\section{Introduction}

 Understanding the behaviour of high energy hadron reactions from a fundamental perspective within of Quantum Chromodynamics (QCD) is an important goal of particle physics (For recent reviews see e.g. Ref. \cite{cgc}). 
{ The central papers concerning the knowledge of the Regge limit (high energy
limit) of perturbative QCD (pQCD) were presented} in the mid seventies by Lipatov and collaborators \cite{BFKL}, which demonstrated that the high energy behaviour of the total cross sections is related to the Pomeron exchange. 
In the framework of perturbative QCD   the Pomeron corresponds to a  $C$-even  parity ($C$ being the charge conjugation) compound state  of two $t$-channel reggeized  gluons, given by the solution of the Balitsky - Fadin - Kuraev - Lipatov (BFKL) equation \cite{BFKL}. 
Besides the Pomeron, a natural prediction of the QCD is the  presence of the so-called Odderon, which is a $C$-odd compound state of three reggeized gluons, which dominates the hadronic cross section difference between the direct and crossed channel processes at very high energies. The Odderon is described by  the Bartels - Kwiecinski - Praszalowicz (BKP) equation \cite{bkp}, which resums terms of the order $\alpha_s(\alpha_s \log s)^n$ with arbitrary $n$ in which three gluons in a $C = -1$ state are exchanged in the $t$-channel.

In the last years, the physics of the Odderon has become an increasingly active subject of research, both from theoretical and experimental  points of view (For a recent review see \cite{ewerz}). On the theoretical side, the investigation of the Odderon in pQCD has led to discovery of relations of high energy QCD to the theory of integrable models \cite{korchemsky} and two leading solutions of the BKP evolution equation were obtained \cite{janik,blv}, with the intercept being close to  or exactly one, depending on the scattering process. In contrast, on the experimental side, 
the existence of the Odderon is still not confirmed. The experimental evidence for the Odderon is at the moment rather scarce. 
A recent study of the data on the differential elastic $pp$ scattering shows that one needs the Odderon to describe the cross sections in the dip region \cite{dosch_ewerz}. The difficulties inherent in the description of $pp$ and $p\bar{p}$ collisions and the lack of  further data have made it impossible to establish the existence of the Odderon in these processes beyond reasonable doubt. However, it is important to emphasize that  the TOTEM collaboration \cite{totem} measured recently the differential cross section for elastic $pp$ collisions at $\sqrt{s}$ 7 TeV and more data are expected in the next years.
An alternative to probe the Odderon is the study of the  diffractive photoproduction of pseudoscalar mesons in $ep$ collisions (For other possibilities see Refs. \cite{pheno_odderon,pheno_odderon2}).
As the real photon emitted by the electron carries negative $C$ parity,  its transformation into a diffractive final state system of positive $C$ parity  requires the $t$-channel exchange of an object of negative $C$ parity. It implies that Pomeron exchange cannot contribute to this process and that it can only be mediated by the exchange of an Odderon. A particular promising process is the diffractive $\eta_c$ photoproduction, since  the meson mass provides a hard scale that makes a perturbative calculation possible \cite{ckms,bbcv}.
Unfortunately, even with an optimistic choice of the strong running coupling constant $\alpha_s$, the cross section for this process is  too small to be observed at HERA.  
Consequently, the current experimental evidence for the Odderon is very unsatisfactory.

\begin{figure}[t]
\centerline{\psfig{file=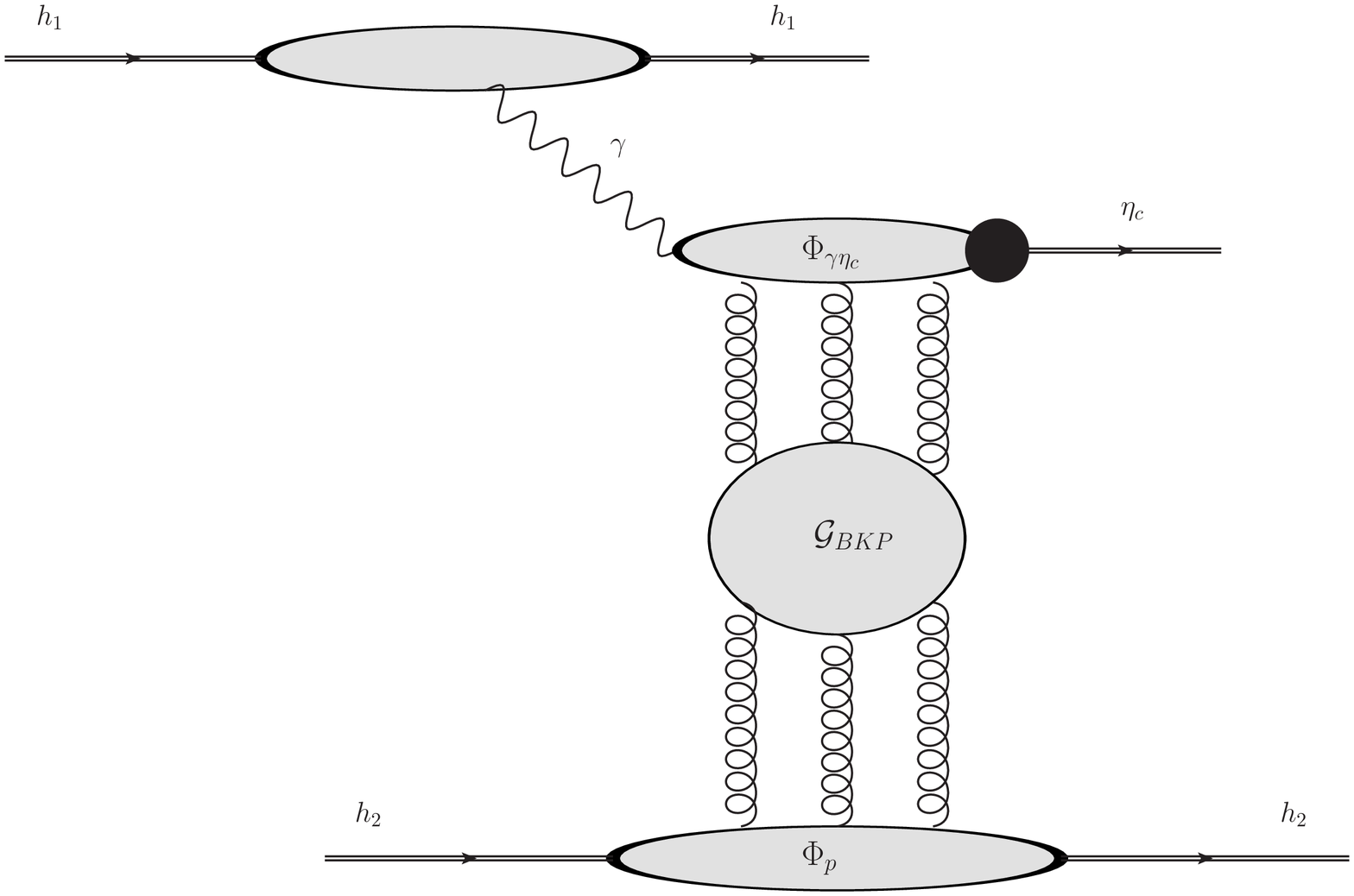,width=80mm}}
 \caption{Illustration of the coherent process $h_1 + h_2 \rightarrow h_1 h_2 \eta_c$ (See text).}
\label{fig1}
\end{figure}

In this paper we propose the study of coherent hadron - hadrons interactions as  a new alternative to search the Odderon in high energy $pp$ and $AA$ collisions at CERN - LHC. In last years several authors  demonstrated that the study of  these interactions is useful to constrain the physics of the Pomeron  at high energies (See, e.g. Refs. \cite{vicmag_upcs,vicmag_update,outros}). 
Here we demonstrate that coherent interactions can also be used to probe the Odderon. 
The basic idea in coherent  hadronic collisions is that the total cross section for a given process can be factorized in
terms of the equivalent flux of photons into the hadron projectile and
the photon-photon or photon-target production cross section. Consequently, coherent interactions can be used to study photon induced processes, as for example, the diffractive $\eta_c$ photoproduction, which is a direct probe of the Odderon. In Fig. \ref{fig1} we present an illustration of the coherent process $h_1 + h_2 \rightarrow h_1 h_2 \eta_c$ which will be analysed in this paper. The main advantage of using colliding hadrons and nuclear beams for studying $\gamma p (A)$ interactions is the high equivalent photon
energies and luminosities that can be { obtained} at existing and
future accelerators (For  reviews see Ref. \cite{upcs}). 

This paper is organized as follows. In Sections \ref{coe} and \ref{dif} we present a  brief  review of  the main concepts and formulas used  in the description of coherent interactions and  diffractive $\eta_c$ photoproduction, which are required  to explain our results, which will presented in Section \ref{results}. Finally, in Section \ref{sum} we summarize our main conclusions.

\section{ Coherent interactions}
\label{coe} 

Lets consider a hadron-hadron interaction at large impact parameter ($b > R_{h_1} + R_{h_2}$) and at ultra relativistic energies. In this regime we expect the dominance of the electromagnetic interaction.
In  heavy ion colliders, the heavy nuclei give rise to strong electromagnetic fields due to the coherent action of all protons in the nucleus, which can interact with each other. In a similar way, it also occurs when considering ultra relativistic  protons in $pp(\bar{p})$ colliders.
The photon stemming from the electromagnetic field
of one of the two colliding hadrons can interact with one photon of
the other hadron (two-photon process) or can interact directly with the other hadron (photon-hadron
process). The total
cross section for a given process can be factorized in terms of the equivalent flux of photons of the hadron projectile and  the photon-photon or photon-target production cross section \cite{upcs}.  In what follows our main focus shall be in photon - hadron processes.
In the particular case of the  diffractive $\eta_c$ photoproduction  in a coherent  hadron-hadron collision, the total cross section is  given by,
\begin{eqnarray}
\sigma (h_1 h_2 \rightarrow h_1 \otimes \eta_c \otimes h_2) =  \sum_{i=1,2} \int dY \frac{d\sigma_i}{dY}\,,
\label{sighh}
\end{eqnarray}
where $\otimes$ represents a rapidity gap in the final state and ${d\sigma_i}/{dY}$ is the rapidity distribution for the photon-target interaction induced by the hadron $h_i$, which can be expressed as 
\begin{eqnarray}
\frac{d\sigma_i}{dY} = \omega \frac{dN_{\gamma/h_i}}{d\omega}\,\sigma_{\gamma h_j \rightarrow \eta_c h_j} (W_{\gamma h_j}^2) \,\,\,\,\,\,(i\neq j)\,.
\label{rapdis}
\end{eqnarray}
$\omega$ is the photon energy,   $\frac{dN_{\gamma/h}}{d\omega}$ is the equivalent photon flux, $W_{\gamma h}^2=2\,\omega \sqrt{s_{\mathrm{NN}}}$  and ${s_{\mathrm{NN}}}$ are  the  c.m.s energy squared of the
photon - hadron and hadron-hadron system, respectively.

Considering the requirement that  photoproduction
is not accompanied by hadronic interaction (ultra-peripheral
collision) an analytic approximation for the equivalent photon flux of a nuclei can be calculated, which is given by \cite{upcs}
\begin{eqnarray}
\frac{dN_{\gamma/A}\,(\omega)}{d\omega}= \frac{2\,Z^2\alpha_{em}}{\pi\,\omega}\, \left[\bar{\eta}\,K_0\,(\bar{\eta})\, K_1\,(\bar{\eta})+ \frac{\bar{\eta}^2}{2}\,{\cal{U}}(\bar{\eta}) \right]\,
\label{fluxint}
\end{eqnarray}
where   $\bar{\eta}=\omega\,(R_{h_1} + R_{h_2})/\gamma_L$ (with $\gamma_L$ being the Lorentz boost  of a single beam),  ${\cal{U}}(\bar{\eta}) = K_1^2\,(\bar{\eta})-  K_0^2\,(\bar{\eta})$
and $K_0(\eta)$ and  $K_1(\eta)$ are the
modified Bessel functions.
The Eq. (\ref{fluxint}) will be used in our calculations of the $\eta_c$  production in $AA$ collisions. On the other hand, for   proton-proton collisions, we assume that the  photon spectrum of a relativistic proton is given by  \cite{Dress},
\begin{eqnarray}
\frac{dN_{\gamma/p}(\omega)}{d\omega} =  \frac{\alpha_{\mathrm{em}}}{2 \pi\, \omega} \left[ 1 + \left(1 -
\frac{2\,\omega}{\sqrt{S_{NN}}}\right)^2 \right] .
\left( \ln{\Omega} - \frac{11}{6} + \frac{3}{\Omega}  - \frac{3}{2 \,\Omega^2} + \frac{1}{3 \,\Omega^3} \right) \,,
\label{eq:photon_spectrum}
\end{eqnarray}
where $\Omega = 1 + [\,(0.71 \,\mathrm{GeV}^2)/Q_{\mathrm{min}}^2\,]$ and $Q_{\mathrm{min}}^2= \omega^2/[\,\gamma_L^2 \,(1-2\,\omega /\sqrt{s_{NN}})\,] \approx (\omega/
\gamma_L)^2$. Considering that in 
$pp/PbPb$ collisions at LHC the Lorentz factor  is
$\gamma_L = 7455/2930$, the maximum center-of-mass energy for the $\gamma h$ system, $W_{\gamma h}$,  will be $ \approx 8390/950$ GeV. Therefore, while studies of photoproduction at HERA were limited to photon-proton center of mass energies of about 200 GeV, photon-hadron interactions at  LHC can reach one order of magnitude higher on energy.
It implies that  studies of $\gamma p (A)$ interactions
at the LHC could provide valuable information on the QCD dynamics.

\section{Diffractive $\eta_c$ photoproduction}
\label{dif}
In what follows we present the cross section for the diffractive $\eta_c$ photoproduction, which is the main input in our calculations [See Eq. (\ref{rapdis})]. It  can be obtained using the impact  factor representation, proposed by Cheng and Wu \cite{ChengWu} many years ago. In this representation, the amplitude for a large-$s$ hard collision process can be factorized in {three parts}: the two impact factors of the colliding particles and the Green's function for the  three interacting reggeized gluons, which is determined by the BKP equation and is  represented by ${\cal G}_\mathrm{BKP}$ hereafter. 
The differential cross section for the process $\gamma + h \rightarrow \eta_c + h$  is given by \cite{bbcv}
\begin{eqnarray}
\frac{d\sigma}{dt} = \frac{1}{32 \pi}\sum_{i=1,2} |{\cal{A}}^i|^2 \,\,, 
\end{eqnarray}
where ${\cal{A}}^i$ is the amplitude for a given transverse polarization $i$ of the photon, which can be expressed as a convolution of the impact factors for the proton ($\Phi_p$) and for the $\gamma \eta_c$ transition ($\Phi^i_{\gamma \eta_c}$) with the Odderon Green function:
\begin{eqnarray}
{\cal{A}}^i = \frac{5}{1152} \frac{1}{(2\pi)^8} \langle \Phi^i_{\gamma \eta_c}|{{\cal{G}}_{BKP}}| \Phi_p \rangle\,\,.
\end{eqnarray}
Differently from $\Phi^i_{\gamma \eta_c}$, that can be calculated perturbatively \cite{ckms}, the impact factor $\Phi_p$ that describes the coupling of the Odderon to the proton is non-perturbative and should be modelled. In our calculations we consider the model used in  Refs. \cite{ckms,bbcv} (We refer the reader to the original papers for the details). 

The Odderon Green function ${\cal{G}}_{BKP}$ is described in terms of the solution of the BKP equation \cite{bkp}, with the energy dependence being determined by the Odderon intercept $\alpha_{\odd}$. 
Currently, two leading solutions of the Odderon evolution equations are available \cite{janik,blv} and the subject continues to be under intensive study (See, e.g., \cite{outros_odderon}). In Ref. \cite{janik}, Janik and Wosiek (JW)  obtained that $\alpha_{\odd} = 1 - 0.24717 \frac{\alpha_s N_c}{\pi}$, which for $\alpha_s \approx 0.2$ yields $\alpha_{\odd} = 0.96$. Moreover, they found that its solution does not couple to all phenomenologically relevant impact factors. In particular, the coupling for the $\gamma  \eta_c$ impact factor vanishes in leading order. In contrast, Bartels, Lipatov and Vacca (BLV) \cite{blv} have found a solution for the BKP equation with intercept $\alpha_{\odd}$ exactly equal to one and that couples  to   $\Phi_{\gamma \eta_c}$, in contrast to the JW solution. Such solution was used in \cite{bbcv} to estimate the diffractive $\eta_c$ photoproduction in the kinematical region probed in $ep$ collisions at HERA (BBCV model hereafter). They find a 
weak logarithmic suppression with the energy and have  predicted an integrated cross section of $\approx$ 50 pb at HERA energy. This prediction is a factor $5$ larger than the value obtained by Kwiecinski and collaborators in Ref. \cite{ckms} (CKMS model hereafter), which has considered a simplified three gluon exchange model for the Odderon that implies an energy independent cross section. A shortcoming of these analysis is the large value  used for the effective strong coupling constant present in the coupling of the Odderon to the external particles. As pointed in Ref. \cite{dosch_ewerz} the cross sections reported in Refs. \cite{bbcv,ckms} have to be reduced by a factor 30. In what follows we consider the approach proposed in 
Ref. \cite{bbcv} and we will use  a more realistic value for $\alpha_s$ (= 0.3). For completeness we also present the results of the approach proposed in \cite{ckms}.

\section{Results}
\label{results}

In Table \ref{I} we present our predictions for the  total cross section  considering  $pp$ collisions at $\sqrt{s_{NN}} = 8$ and 14 TeV. We predict values of the order of $pb$, with the BBVC prediction being a factor of $\approx 20$ larger than the CKMS one. This large factor of enhancement is directly associated to the energy dependence present in the BBVC model, which implies that the $\gamma h$ cross section increases at smaller energies, while the CKMS predicts an energy independent cross section. It is important to emphasize that the main contribution for the total $h_1h_2$ cross section comes from  small values of $\omega$ due to dependence of the equivalent photon spectrum in the photon energy, which is proportional to $1/\omega$. Furthermore, the increasing of $dN/d\omega$  with $\sqrt{s_{NN}}$ implies that $\sigma      
(h_1 h_2 \rightarrow h_1 \otimes \eta_c \otimes h_2)$ also increases with the center-of-mass energy.
In comparison to the cross sections for the $J/\Psi$ photoproduction \cite{vicmag_update}, our predictions are a factor $\ge 10^3$ smaller, with the difference increasing with the energy due to the Pomeron exchange present in the $J/\Psi$ production. On the other hand, in comparison to the predictions for exclusive vector meson production from the pomeron - odderon fusion presented in \cite{pheno_odderon2}, our results are similar to those for the $\Upsilon$ hadroproduction.

In Table \ref{II} we present our predictions for the  total cross section considering  $PbPb$ collisions at two values of center-of-mass energy. 
As the photon flux is proportional to $Z^2$,
{because} the electromagnetic field surrounding the ion is very larger than the proton one due to the coherent action of all protons in the nucleus,  the nuclear cross sections are amplified by this factor. Moreover, our predictions also are amplified by the mass number $A$, since in  our calculations for the nuclear case we are 
assuming in a first approximation that $\sigma (\gamma A \rightarrow \eta_c A) = A . \sigma (\gamma p \rightarrow \eta_c p)$. Consequently, we predict   cross sections of the order of $\mu b$ for the diffractive $\eta_c$ photoproduction in $PbPb$ collisions at LHC.

Considering the {design} luminosities at LHC for $pp$ collisions (${\cal L}_{\mathrm{pp}} = 10^{34}$ cm$^{-2}$s$^{-1}$) and $PbPb$ collisions (${\cal L}_{\mathrm{PbPb}} = 4.2 \times 10^{26}$ cm$^{-2}$s$^{-1}$) we can calculate the production rates (See Tables \ref{I} and \ref{II}). Although the cross section for the diffractive $\eta_c$ photoproduction  in $AA$ collisions is much larger than in $pp$ collisions, the event rates are higher in the $pp$ mode  due to its larger luminosity. In particular, we predict that the events rate/year for $pp$ collisions at $\sqrt{s} = 14$ TeV should be larger than 65000. However, as emphasized in Ref. \cite{david}, for a luminosity above ${\cal L} \ge 10^{33}$ cm$^{-2}$s$^{-1}$, multiple hadron - hadron collisions per bunch crossing are { very likely}, which leads to a relatively large occupancy of the detector channels even at low luminosities.
{This} drastically reduces the possibility of {measurement} of coherent
processes at these luminosities. In contrast, for  lower luminosities, the event pile-up is negligible. Consequently, an estimative considering ${\cal L}_{\mathrm{pp}} = 10^{32}$ cm$^{-2}$s$^{-1}$ should be more realistic. It reduces our predictions for the event rates in $pp$ collisions by a factor $10^2$.

\begin{table}[t]
\begin{center}
\begin{tabular}{||c|c|c||}
\hline 
$\sqrt{s_{NN}}$ & CKMS & BBCV \\
\hline
\hline 
8 TeV & $0.55\, pb$ (55000) & $10.10\, pb$ ($1 \times 10^6$) \tabularnewline
\hline 
14 TeV & $0.65\, pb$ (65000) & $13.90\, pb$ ($1.4 \times 10^6$) \tabularnewline
\hline
\end{tabular}
\caption{Cross sections (event rates/year) for the diffractive $\eta_c$ photoproduction in $pp$ collisions at LHC energies. }
\label{I}
\end{center}
\end{table}

\begin{table}[t]
\begin{center}
\begin{tabular}{||c|c|c||}
\hline 
$\sqrt{s_{NN}}$ & CKMS & BBCV \\
\hline
\hline 
2.76 TeV & $0.30 \, \mu b$ (126) & $14.25 \, \mu b$ (5985) \tabularnewline
\hline 
5.5 TeV & $0.40 \, \mu b$ (168) & $23.59 \, \mu b$ (9912)\tabularnewline
\hline
\end{tabular}
\caption{Cross sections (event rates/year) for the diffractive $\eta_c$ photoproduction in $PbPb$ collisions at LHC energies. }
\label{II}
\end{center}
\end{table}

A final comment about the experimental separation of the $\eta_c$ produced in a photon - hadron interaction at LHC is in order. 
The Odderon and the photon exchanged in Fig. \ref{fig1} are colorless objects, which lead to the formation of two rapidity gaps in the final state, i.e. the outgoing particles ($h_1$, $\eta_c$ and $h_2$) are separated by a large region in rapidity in which there is no hadronic activity observed in the detector. 
Two rapidity gaps in the final state also are generated in photon - photon interactions. The associated cross sections for production of $\eta_c$ in these interactions for $pp$ \cite{daniel} and $PbPb$ \cite{bertulani} collisions are almost a factor six larger than those presented in this paper. Consequently,  the detection of two rapidity gaps in the final state is not, in a first analysis, an efficient trigger for the separation of the $\gamma h$ production of the $\eta_c$. As already emphasized, the use of this trigger at LHC will be a hard task due to the  multiple hadron - hadron collisions per bunch crossing, which are not negligible at high luminosities. Simulations indicate that coherent processes could be identified with a good signal to background ratio when the entire event is reconstructed and a cut is applied on the summed transverse momentum of the event \cite{david}. 
As the typical photon virtualities are very small, the hadron scattering angles are very low. Consequently, we expect that  
a different transverse momentum distribution of the scattered hadrons, with $\gamma h$ interactions predicting larger $p_T$ values. Therefore,
we believe that the detection of the scattered hadrons with extra detectors in a region away from the interaction point probably can eliminate many serious backgrounds and to probe the Odderon. Currently, almost all LHC collaborations have a program of forward physics. Preliminary results indicate that  the study of coherent processes at LHC and its experimental separation is feasible  (See, e.g. \cite{forward,david2}). However, this subject  certainly deserves more detailed studies.

\section{Summary}
\label{sum}
Despite the enormous success of QCD there remains a number of deep questions to be answered in the field of strong interaction physics. In particular, the existence of the Odderon, which is a unambiguous prediction of QCD,  is still not confirmed in the experiment. In this paper we  proposed the probe of the perturbative Odderon in coherent hadron - hadron interactions at LHC. We study the diffractive $\eta_c$ photoproduction, which is an exclusive process in which the 
Odderon is the only possible exchange. Hence the observation of such processes would clearly indicate the existence of the Odderon. We predict  total cross sections of order of $pb \, (\mu b)$ for $pp \, (PbPb)$ collisions and large values for the event rates/year, which makes, in principle, the experimental analysis of this process feasible at LHC. Our results are complementary to the predictions presented in Ref.  \cite{pheno_odderon2}, where the probe of the perturbative Odderon  in the exclusive vector meson hadroproduction was proposed. As in   \cite{pheno_odderon2}, the detection of the outgoing protons will be necessary in order to eliminate the main backgrounds and to probe the Odderon. 

Finally, we point out that another alternative to probe the Odderon is the study of the inclusive $\eta_c$ production, described at the photon level by the process $\gamma + h \rightarrow \eta_c +X$ in the triple Regge region, where $X$ is a diffractive system, which contains the coupling of the Pomeron to  two Odderons \cite{bbcv2}. The cross section for this process is expected to be much larger than the diffractive one \cite{future}.



.

\section*{Acknowledgements}
 This work was partially financed by the Brazilian funding agencies CNPq and FAPERGS.



\end{document}